\documentclass[11pt,fleqn]{article}   
\usepackage{latexsym}
\catcode`\@=11
\@addtoreset{equation}{section}
\def\theequation{\arabic{section}.\arabic{equation}}
\def\appendix{\renewcommand{\thesection}{\Alph{section}}\setcounter{section}{0}
              \renewcommand{\theequation}
            {\mbox{\Alph{section}.\arabic{equation}}}\setcounter{equation}{0}}
\def\maketitle{\thispagestyle{empty}\setcounter{page}0\newpage
                \renewcommand{\thefootnote}{\arabic{footnote}}
                  \setcounter{footnote}0}
\renewcommand{\thanks}[1]{\renewcommand{\thefootnote}{\fnsymbol{footnote}}
               \footnote{#1}\renewcommand{\thefootnote}{\arabic{footnote}}}

\renewcommand{\title}[1]{\begin{center}\Large\bf #1\end{center}\rm\par\bigskip}
\renewcommand{\author}[1]{\begin{center}\Large #1\end{center}}
\newcommand{\address}[1]{\begin{center}\large #1\end{center}}

\def\dinfn{\smallskip Dipartimento di Fisica, Universit\`a di Trento\\ 
                           and Istituto Nazionale di Fisica Nucleare,\\
                                   Gruppo Collegato di Trento, Italia}

\def\Idinfn{\address{\dinfn}}

\newcommand{\email}[1]{e-mail: \sl #1@science.unitn.it\rm}

\newcommand{\femail}[1]{\thanks{\email{#1}}}

\newcommand{\pacs}[1]{\smallskip\noindent{\sl PACS numbers:
                       \hspace{0.3cm}#1}\par\bigskip\rm}
\def\babs{\hrule\par\begin{description}\item{Abstract: }\it} 
\def\eabs{\par\end{description}\hrule\par\medskip\rm}
\renewcommand{\date}[1]{\par\bigskip\par\sl\hfill #1\par\medskip\par\rm}

\newcommand{\s}[1]{\section{#1}}

\newcommand{\ca}[1]{{\cal #1}}         
\def\hs{\qquad}               
\def\nn{\nonumber}            
\def\beq{\begin{eqnarray}}    
\def\eeq{\end{eqnarray}}      
\def\ap{\left.}               
\def\at{\left(}               
\def\cp{\right.}              
\def\ct{\right)}              
\def\R{{\hbox{{\rm I}\kern-.2em\hbox{\rm R}}}}   
\def\H{{\hbox{{\rm I}\kern-.2em\hbox{\rm H}}}}   
\def\N{{\hbox{{\rm I}\kern-.2em\hbox{\rm N}}}}   
\def\C{{\ \hbox{{\rm I}\kern-.6em\hbox{\bf C}}}} 
\def\Z{{\hbox{{\rm Z}\kern-.4em\hbox{\rm Z}}}}   
\def\ii{\infty}                                  

\def\be{\beta}

\def\de{\delta}

\def\ka{\kappa}
\def\la{\lambda}

\def\si{\sigma}

\begin{document}

\title{Action, Hamiltonian and CFT for $2D$ black holes}

\author{M.~Caldarelli\femail{caldarel}, G.~Catelani and L.~Vanzo\femail{vanzo}}

\Idinfn

\begin{abstract}
The boundary terms in the Hamiltonian, in the presence of horizons, are carefully
analyzed in a simple $2D$ theory admitting AdS black holes. The agreement
between the Euclidean, Gibbons-Hawking approach and CFT through Cardy's formula
is obtained modulo certain assumptions regarding the spectrum of the Virasoro's
algebra. There is no discrepancy factor $\sqrt{2}$ once
the appropriate boundary conditions are properly recognized. The results 
obtained here are of general validity, since they rely on general properties of 
black holes. In particular, the central charge can be understood as a
classical result without invoking a string or any other microscopic
theory. The peculiar feature of gravity, that the on-shell Hamiltonian
is determined by boundary terms, is the reason of the mentioned
agreement. 
\end{abstract}


\s{Introduction}

Gravity in $2D$ is a very rich and much investigated subject (a review of early work
is in the TASI lectures by Ginsparg-Moore\cite{tasi93}). Under the spell of modern
string theory, there has been recently a lot of activity aiming to
understood the subtle aspects of AdS/CFT correspondence in $2D$
gravity\cite{stro1,stro2,gibb,mmst,cami,ckz,caca} (even two
dimensional de Sitter space has been understood as a variant of AdS/CFT
correspondence\cite{hms00}). \\
A consistent piece of work in $2D$ gravity has always been to understand $AdS_2$
black holes\cite{bht86,lemo96,andy99,cado00,cksz00}. Most researches almost invariably
focused on the asymptotic symmetries of $AdS_2$ gravitating system. It seems
instructive, and to some extent important, to look instead at the near horizon
symmetries, and the purpose of this letter is to do this in the Hamiltonian
framework of AdS gravity (see \cite{nayo99} for applications to AdS/CFT
correspondence). We first determine the regularity (as opposed to boundary)
conditions near the horizon. We then proceed to find the corresponding
Hamiltonian. A new term at the horizon will be found, which can be interpreted
as a shift in energy due to a central charge in the diffeomorphism algebra. This
will be used to evaluate the entropy, which will then be compared with the standard
thermodynamics of black holes, finding complete agreement (no $\sqrt{2}$
discrepancy\cite{cado00}).    

\s{Action and Hamiltonian}

The theory we will consider is the one defined by the
Jackiw-Teitelboim's action\cite{jtei}  
\beq
I=\frac{1}{2}\int|g|^{1/2}\eta(R+2\la^2)
\eeq
which leads to the equations of motion
\beq
R+2\la^2=0, \hs g_{ab}\Box\eta-\nabla_a\nabla_b\eta=\la^2g_{ab}\eta 
\eeq
Black holes in this model have been studied in a number of
papers\cite{bht86,lemo96,cksz00,cado99,nana99}. The standard black hole
solution is the linear dilaton vacuum, with metric
\beq
ds^2=-(\la^2x^2-a^2)dt^2+(\la^2x^2-a^2)^{-1}dx^2
\label{metr}
\eeq
and dilaton $\eta=\eta_0\la x$, where $a$ and $\eta_0$ are integration
constants. The event horizon is located at $x=x_0=\la^{-1}a$. 
Given that the solution is uniquely determined by (i),
existence of a timelike $U(1)$ isometry and (ii), asimptotically AdS
behaviour, one may wonder why the black hole should exhibit any
thermodynamics property at all. \\
To better understanding this point, 
we are going to discuss how big is the phase space of the black hole,
and to this aim we need the Hamiltonian. We start then with a general metric
\beq
ds^2=-N^2dt^2+\si^2(dx+Vdt)^2, \hs |g|^{1/2}=N\si
\eeq
and define $U=-\dot{\si}+(\si V)^{'}$, where a prime means derivative w.r.to $x$. 
A calculation shows that
\beq
I&=&I({\rm bulk})-\lim_{X\to\ii}\int
dt\,\eta(\si^{-1}N^{'}-N^{-1}VU)-\left[\int_{x_0}^X dx\eta
N^{-1}U\right]^{t_1}_{t_2} \nn \\
&+&\lim_{x\to x_0}\int dt\left(\eta\si^{-1}N^{'}-N^{-1}\eta VU\right)
\eeq
with 
\beq
I(bulk)&=&-\int_{x_0}^X\left(\frac{\dot{\eta}\dot{\si}}{N}-\frac{\eta^{'}N^{'}}{\si}
+\frac{V(V\si)^{'}\eta^{'}}{N}-\frac{\dot{\si}V\eta^{'}}{N}\cp\nn \\
&&\ap-\frac{\dot{\eta}(\si V)^{'}}{N}-\la^2N\si\eta\right)d^2x
\eeq
Of course, in doing so one is concerned with the evolution in the exterior
region, marked by $X>x>x_0=a\la^{-1}$, where $X$ will be taken to
infinity when necessary. To simplify things let us set
$V=0$ from now on. We also omit the limits of spatial integration, 
leaving it understood that the range is actually $x\in[x_0,X]$. Under a
small change of the metric and dilaton, the action has a variation
\beq
\de I&=&{\rm ``terms\,\, giving\,\, equations\,\, of\,\, motion}''\,\,-
\int dx\left[N^{-1}\dot{\eta}\de\si\right]^2_1 \nn \\ 
&+&\int dt\left[\si^{-1}\eta^{'}\de N\right]^X_{x_0}-\int dt
\left[\eta\de\left(\frac{N^{'}}{\si}\right)\right]^X_{x_0}+\int dx\left[\eta
\de\left(\frac{\dot{\si}}{N}\right)\right]^2_1
\label{vari}
\eeq
Before going any further, a crucial issue will be to decide what kind of
regularity conditions have to be
imposed near the black hole horizon (see \cite{wiyo88} for a discussion in
four dimensions). This is the place where the $U(1)$
isometry of time translation has a fixed point. We also require that
near the horizon the geometry be isometric to a flat disk, so that the
Euler characteristic will be $\chi=1$. Both conditions can be met
by\footnote{They were also discussed in thermodynamics ensembles by
J. D. Brown and collaborators\cite{bcmmwy90}.}
\beq
N(x_0)=0, \hs  (\si^{-1}N^{'})_{|x=x_0}=\ka
\label{regu}
\eeq
where $\ka$ is the surface gravity of the black hole. We also needs to fix
the metric, the lapse and $\eta$ at $t_1$, $t_2$ and infinity\footnote{Because 
we are going to discuss the canonical ensemble.}. Looking at
(\ref{vari}), we see that we have to refine the action to read
\beq
I_{NEW}=I+\int dt\,(\si^{-1}\eta N^{'})_{|X}-\int
dx\,\left[N^{-1}\eta\dot{\si}\right]^2_1
\eeq
Even without bothering about boundary terms, we can write a canonical
Hamiltonian associated with $I$\footnote{A rather general treatment of
phase space formulations of $2D$ gravity models can be found in
\cite{krv97}, and references therein.}. After standard manipulations, we obtain
\beq
H&=&\int dx\,N\left[-P_{\si}P_{\eta}+\left(\frac{\eta^{'}}{\si}\right)^{'}
-\la^2\si\eta\right]-\left(\frac{\eta^{'}N}{\si}\right)_{|X} \nn \\
&+&\at\frac{\eta^{'}N}{\si}\ct_{|x_0}+\at\frac{\eta N^{'}}{\si}\ct_{|X}-
\at\frac{\eta N^{'}}{\si}\ct_{|x_0}
\label{hami}
\eeq
where $P_{\si}=-N^{-1}\dot{\eta}$, $P_{\eta}=-N^{-1}\dot{\si}$ and $P_N=0$. The
first two boundary terms have appeared because we wrote the Hamiltonian in a 
form that displays the constraint
\beq
\ca H_{\perp}=-P_{\si}P_{\eta}+\left(\frac{\eta^{'}}{\si}\right)^{'}
-\la^2\si\eta
\eeq
The last term in (\ref{hami}) is the crucial one, that will be related
to a Virasoro's central charge. It is clear that it exists only
because we wanted to predict the outer region of the black hole. The
Hamiltonian on a geodesically complete spatial section of the black
hole would not have any horizon term. In this respect it makes sense to
interpret it as a sort of classical entanglement of states. It is a
very peculiar feature of gravity, because only gravity has the
Hamiltonian which on-shell is given by pure boundary terms.\\ 
The variation of $H$ becomes
\beq
\de H=-\left[\si^{-1}\eta^{'}\de N-\si^{-1}\eta\de N^{'}+\si^{-2}\eta N^{'}
\de\si\right]^{X}_{x_0}
\eeq
plus bulk terms giving the equations of motion. So again we should
define 
\beq
H_{NEW}=H-\at\si^{-1}\eta N^{'}\ct_{|X}
\eeq
and this will be suitable to our boundary conditions. Notice that the $\de\eta$
never appears on the horizon, so we need not a boundary condition for $\eta$ at
$x_0$. Finally
\beq
H_{NEW}&=&{\rm constraint\,\,term}\,\,-\at\si^{-1}\eta
N^{'}\ct_{|x_0}+N(X)[(\si^{-1}\eta^{'})_{|B}-(\si^{-1}\eta^{'})_{|X}] \nn  \\
&+&\at\si^{-1}\eta^{'}N\ct_{|x_0}
\eeq
the last term being really zero for a dilaton regular on the horizon. We also
included a background subtraction, labelled by the subscript $''B''$,
to insure the convergence of the
limits as $X\to\ii$. The natural ground state would be in the same topology class
as the black hole and would have zero temperature. This identifies it as the metric
(\ref{metr}) with $a=0$. The horizon boundary term in $H_{NEW}$ is an instance
of a general result pertaining to spacetimes with degenerate
foliations\cite{bmy91,hh99}.\\ 
We are now in position to describe the (reduced) phase space of the
black hole (\ref{metr}). With a linear dilaton, the constraint implies
$\si^2=(\la^2x^2-a^2)^{-1}$, but leaves undetermined the
lapse and sets the momentum variables to zero. 
However, the horizon regularity conditions (\ref{regu}) fix
the behaviour of $N$ near the horizon to have the form
\beq
N^2=\frac{2\ka^2}{\la a}(x-a/\la)+\be(t)(x-a/\la)^2+O_3
\label{expa}
\eeq
and near infinity to be of order $\la^2x^2$, but leaves $N$ otherwise
arbitrary. So, roughly speaking, there are as many metrics with a
given surface gravity as there are analytic functions on the
disk, vanishing in the origin and satisfying (\ref{expa}). Certainly
this includes the full conformal  group in $2D$, as this also is
generated by analytic functions on the disk. 

\s{Thermodynamics and CFT}

The boundary term at infinity in $H_{NEW}$ can be easily evaluated,
and leads to identify the mass as
\beq
M=\frac{1}{2}\eta_0\la a^2
\eeq
The boundary term on the bifurcation point of the horizon is new, and
has the effect to shift the mass to a lower value. 
In the spirit of CFT, we interpret then the Hamiltonian as
\beq
H_{NEW}=\ka\left(L_0+\tilde{L}_0-\frac{c+\tilde{c}}{24}\right)
\label{cfth}
\eeq
The other Virasoro's generators $L_n$, $\tilde{L}_n$, would be just
the on shell value of 
the Hamiltonians $H[\xi_n]$, $H[\tilde{\xi}_n]$ for surface
deformations\cite{teit73} generated by vector fields respecting the horizon
regularity conditions near the bifurcation point\cite{vanz00}. We may 
use the time inversion symmetry of the solution to infer that, in fact, 
$c=\tilde{c}$. This stems from the fact that static black holes do have 
indeed a pair of classical Virasoro algebras connected with the canonical 
realizations of symmetries on the phase space\cite{carl99,solo99}, but these 
are linked to the future and past sheet of the horizon, respectively\cite{vanz00}.\\ 
Since the horizon term in $H_{NEW}$ is $-\ka\eta_{hor}=-\ka\eta_0a$,  this gives 
a central charge
\beq
c=12\eta_0a
\eeq
What's about $L_0$ and $\tilde{L}_0$? In a Lorentzian world there are in fact two
horizons, but in the canonical Euclidean picture there is only one, the axis of
the Euclidean section. It is natural to consider the sector of the CFT in which
all Virasoro generators of the past horizon, $\tilde{L}_n$, vanish for
$n\leq 0$\footnote{The generators $\tilde{L}_n$, $n>0$ vanish too,
because they correspond to spacetime diffeomorphisms having a zero of
order $n$ in $(x-x_0)$\cite{vanz00}.}. This means imposing that white holes
have no entropy. But then $L_0=\eta_0a/2$ and Cardy's formula gives the
entropy as
\beq
S=2\pi\sqrt{\frac{cL_0}{6}}=2\pi\eta_0a
\eeq
We may confirm this simply comparing $T_HdS$ with $dM$, where $T_H=(2\pi)^{-1}\la
a$ is the Hawking's temperature of the black hole. Or we may 
evaluate the off-shell Euclidean action (a la
Gibbons-Hawking\cite{gibb77}), which is of independent interest,
choosing as a reference background the zero temperature state (the
extreme black hole metric with $a=0$). Then we obtain
\beq
I_E=-\frac{1}{2}\int|g|^{1/2}\eta(R+2\la^2)
+\lim_{X\to\ii}\int_0^{\be}d\tau\,[N\eta_B N^{'}_B-\si^{-1}\eta N^{'}]_{|X}
\eeq
and we have also included the background subtraction. We recall that this has
the same dilaton $\eta_B=\eta_0\la x$ and $N_B=\la x$. Using the
Gauss-Bonnet theorem for a disk we obtain 
\beq
I_E=-\frac{(2\pi)^2\eta_0}{\la\be_H}+\frac{(2\pi)^2\eta_0}{2\la\be_H^2}\,\be
\eeq
where $\be_H=2\pi\la^{-1}a^{-1}$ is the Hawking's inverse temperature of
the black hole (\ref{metr}) and $\be$ is arbitrary. Since $\log Z=-I_E$, this 
leads to a mass $M=2^{-1}\eta_0\la a^2$ and entropy
$S=2\pi\eta_0a$, as expected from the Hamiltonian and the black hole
first law, respectively. \\
The on shell action is $I_E$ for $\be=\be_H$, that is
\beq
I_E\equiv-\log Z=-\frac{(2\pi)^2\eta_0}{2\la}\be_H^{-1}
\eeq
This is the tipical behaviour of the partition function of a scale
invariant theory in $2D$, and again gives $M$ and $S$ in agreement
with Cardy's formula, solving also the $\sqrt{2}$ puzzle. \\
The partition function of AdS black holes in
higher dimensions also has the high temperature behaviour of scale
invariant theories, $\log Z\sim
C\be^{-(D-1)}$ \cite{hp83,va97,blp97,wit98,bir99}, but this is now
understood as a consequence of duality with a gauge
theory\cite{mal98}. In $AdS_2$ the question of the dual field theory
is harder, although recently a definite proposal has been
made\cite{caca}, which also solves the $\sqrt{2}$ puzzle. We do not
rely, however, on microscopic details. The more puzzling remains 
the Schwarzschild black hole, which has $\log Z\sim\be^2$, not that
of a scale invariant theory, but that nevertheless has a horizon
central charge\footnote{Since the horizon boundary term in the
Hamiltonian must be present.} matching its entropy, if one uses Cardy's
formula with the same assumptions made here on the Virasoro spectrum.

\s{Conclusion}

It is certainly amusing that thermodynamics and Cardy's formula
match each other to such a high degree. What we find even more
remarkable is that a knowledge of the classical Hamiltonian is
sufficient to determine the central charge of a quantum Virasoro
algebra without appealing to any symmetry consideration. This result
is not limited to two dimensional models and will be further discussed
in a separate paper. Due to these facts, we have confidence that our
assumptions on the spectrum of the Virasoro algebra correctly describe
$2D$ black holes. It also implies that white holes have no
entropy. That the central charge appears in the classical hamiltonian
for the outer region of the black hole, while it is absent for a complete
Cauchy surface, can be interpreted as the classical limit of a sort
of entanglement of states.

\end{document}